\documentclass[twocolumn,floatfix,prb,aps,showpacs]{revtex4-2}
\usepackage{graphicx,amsmath,amssymb,color}
\usepackage{nicefrac}
\usepackage[titletoc,title]{appendix}
\usepackage{float}
\usepackage{orcidlink}

\usepackage{hyperref}
\hypersetup{colorlinks,bookmarks=true,citecolor=blue,linkcolor=red,urlcolor=blue}

\begin{document}

\title{Prediction of non-Abelian fractional quantum Hall effect at $\nu{=}2{+}\frac{4}{11}$ }
\author{Koyena Bose and Ajit C. Balram\orcidlink{0000-0002-8087-6015}}
\affiliation{Institute of Mathematical Sciences, CIT Campus, Chennai, 600113, India}
\affiliation{Homi Bhabha National Institute, Training School Complex, Anushaktinagar, Mumbai 400094, India}
\date{\today}
	
\begin{abstract}
		
The fractional quantum Hall effect (FQHE) in the second Landau level (SLL) likely stabilizes non-Abelian topological orders. Recently, a parton sequence has been proposed to capture many of the fractions observed in the SLL [Ajit C. Balram, 
SciPost Phys. {\bf 10}, 083 (2021)]. We consider the first member of this sequence which has not yet been studied, which is a non-Abelian state that occurs at $4/11$. As yet FQHE in the SLL at this fraction has not been observed in experiments. Nevertheless, by studying its competition with other candidate FQHE states in the SLL we show that this parton state might be viable. We also make predictions for experimentally measurable properties of the parton state which can distinguish it from other topological orders. 
\end{abstract}	
\maketitle

\section{Introduction}
The discovery of the integer quantum Hall effect (IQHE) by von Klitzing, Dorda, and Pepper~\cite{Klitzing80} showed that electrons confined to two dimensions under high magnetic fields and at very low temperatures can exhibit quantization in macroscopic quantities like the Hall resistance. The integral quantization of the transverse resistance is well understood as arising from single-particle Landau level (LL) quantization under a perpendicular magnetic field and Anderson-localization due to the ubiquitous disorder. The field of quantum Hall effect took a new turn with the discovery of fractional quantization of Hall resistance at filling factor $\nu{=}1/3$, known as the fractional quantum Hall effect~\cite{Tsui82}. An understanding of the phenomenon of FQHE necessitates incorporating the Coulomb interaction between the electrons which lifts the single-particle degeneracy within a LL and gives rise to incompressible states at certain rational filling factors. A solution to the $1/3$ FQHE was given by Laughlin~\cite{Laughlin83} who proposed a many-body wave function for the ground state, that was subsequently shown to have an excellent overlap with the exact Coulomb ground state in the lowest Landau level (LLL)~\cite{Ambrumenil88, Kusmierz18, Balram20b}. After that, many fractions of the form $n/(2pn{\pm} 1)$, with $n,p$ positive integers, were observed in the LLL, and these lie beyond the purview of Laughlin's theory. FQHE in the LLL at these fractions can be understood using the composite fermion (CF) theory proposed by Jain~\cite{Jain89}. The CF theory converts an interacting problem of electrons to a noninteracting problem of composite fermions, which are bound states of electrons and an even number of vortices. The FQHE state of electrons at $\nu{=}n/(2pn {\pm} 1)$ corresponds to the IQHE state of CFs carrying $2p$ vortices and filling $n$ of their Landau-like levels.

Over the years as the sample quality has improved, fractions that lie beyond the paradigm of noninteracting CFs have been observed. These fractions have predominantly been observed in the second LL starting with the observation of the famous 5/2 plateau~\cite{Willett87}. Although the 5/2 state can be understood as a topological $p$-wave paired state of CFs which arises from a residual attractive interaction between them~\cite{Moore91, Read00}, generically, it is hard to understand the observed states using CFs. One way to understand these states is using the parton theory~\cite{Jain89b}, which generalizes the CF theory, to map the FQHE state of electrons to IQHE states of partons that are fermionic particles carrying fractional charges. Recently, a parton sequence that captures most of the fractions observed in the SLL has been proposed~\cite{Balram18, Balram19, Balram20a}. In this work, we look at the next unexplored member of this parton sequence which produces a non-Abelian state at $\nu{=}4/11$. As yet, no FQHE has been observed at this fraction in the SLL, though with improving sample quality~\cite{Chung21} incompressibility might arise at $2{+}4/11$. 

In this work, we study the competition in the SLL between the non-Abelian parton state, the non-Abelian Bonderson-Slingerland state~\cite{Bonderson08}, and an Abelian parton state~\cite{Balram21a} that all occur at $4/11$. Our results suggest that all three candidate states are in close competition with each other and only experiments can tell them apart. To this end, we discuss the underlying topological order of these states and make predictions for many measurable quantities that can be tested out in experiments.  

The article is organized as follows: In Sec.~\ref{sec: parton}, we introduce the parton theory, discuss the parton sequence that has been previously proposed for the SLL (Sec.~\ref{subsec: parton_SLL}), and use that as a motivation to put forth a trial wave function for the $2{+}4/11$ FQHE (Sec.~\ref{subsec: parton_4_11}). In Sec.~\ref{sec: results}, we present the results of per-particle SLL Coulomb energies and pair-correlation functions for all three candidate states that we have considered. To obtain the SLL Coulomb energies, we construct a new effective interaction (Sec.~\ref{subsec: eff_int}) that improves upon the effective potentials that have been constructed previously. We conclude the paper in Sec.~\ref{sec: discussion} with a discussion of experimentally measurable properties of the candidate states which could help identify underlying topological order at $2{+}4/11$. 

\section{Parton theory}
\label{sec: parton}

The parton theory proposes that the FQHE states of interacting electrons can be constructed from IQHE states of noninteracting particles called partons~\cite{Jain89b}. Each electron is decomposed into an odd $k$ species of partons, labeled by $(\lambda{=}1,2,...,k)$, which are placed into IQHE states at filling factors $\{ n_{\lambda} \}$. The density of partons is identical to that of the electrons and the partons are exposed to the same external magnetic field as the electrons. This enforces the charge of the $\lambda$ parton species to be $q_{\lambda}{=}\nu ({-}e)/n_{\lambda}$, where ${-}e$ is the charge of the electron. The constraint that the sum of the charges of the partons should add up to that of the electron leads to the relation $\nu{=}(\sum_{\lambda} n_{\lambda}^{{-}1})^{{-}1}$. 

The many-body wave function of the partonic FQHE state denoted by ``$n_{1}...n_{k}$" is~\cite{Jain89b}
\begin{equation}
\Psi^{n_{1} \cdots n_{k}}_{\nu}=\mathcal{P}_{\rm LLL} \prod_{\lambda=1}^k \Phi_{n_{\lambda}},
\label{eq: parton}
\end{equation}
where $\Phi_{n}$ is the IQHE state of $n$ filled LLs of electrons (with $\Phi_{{-}n}{\equiv}\Phi_{\bar{n}}{=}[\Phi_{n}]^{*}$) and $\mathcal{P}_{\rm LLL}$ projects the state to the LLL as is appropriate for the high magnetic field limit of our interest. When the filling factor $n_{\lambda}{<}0$, the corresponding parton carries a positive charge and senses an effective magnetic field that is opposite in direction to that seen by the electrons. The Wen-Zee shift~\cite{Wen92}, which is used to characterize the topological order of a state, for the wave function given in Eq.~\eqref{eq: parton} is $\mathcal{S}{=}\sum_{\lambda{=}1}^{k} n_{\lambda}$. Recent numerical calculations suggest that the projected and unprojected parton states are likely to describe the same topological phase~\cite{Anand22}. Assuming this to be the case, we can work out the topological properties of the unprojected parton states from their Chern-Simons field-theoretic description~\cite{Wen91}. 

Many well-known Abelian FQHE states can be obtained from the parton construction. The $1/p$ Laughlin state~\cite{Laughlin83} is an $11{\cdots}{\equiv}1^{p}$ parton state described by the wave function $\Psi^{\rm Laughlin}_{1/p}{=}\Phi_1^p$. The $n/(2pn {\pm} 1)$ Jain state~\cite{Jain89} is an ${\pm n}11{\cdots}$ parton state described by the wave function $\Psi^{\rm Jain}_{n/(2pn {\pm} 1)}{=} \mathcal{P}_{\rm LLL} \Phi_{{\pm} n} \Phi_1^{2p}$. The parton theory can also produce non-Abelian states~\cite{Wen91} when a repeated factor of $\Phi_n$ with $|n|{\geq 2}$ is present in the wave function given in Eq.~\eqref{eq: parton} and these states can potentially have applications in fault-tolerant topological quantum computation~\cite{freedman2003topological, Nayak08}. Although our focus here would be on the SLL, we mention here that viable candidate parton states, that lie beyond those described by free CFs, have been constructed for FQHE observed in graphene~\cite{Wu17, Kim19, Faugno20a, Faugno21, Balram21b, Dora22, Sharma22}, in wide quantum wells~\cite{Faugno19} and at delicate fractions in the LLL~\cite{Balram21a, Balram21c}.

\subsection{Parton sequence for the second Landau level}
\label{subsec: parton_SLL}
Recently, it has been proposed that the $[\bar{2}1]^{p}\bar{n}1^{2}$ parton sequence can capture many of the experimentally observed SLL states. The states in this parton sequence are described by the wave functions~\cite{Balram18, Balram19, Balram20a}
\begin{equation}
\Psi^{[\bar{2}1]^{p}\bar{n}1^{2}}_{\frac{2n}{(p+4)n-2}} = \mathcal{P}_{\rm LLL} [\Phi_{-2}\Phi_{1}]^{p}~\Phi_{-n}\Phi^{2}_{1} \sim\left[\frac{\Psi^{\rm Jain}_{2/3}}{\Phi_{1}} \right]^{p} \Psi^{\rm Jain}_{n/(2n-1)}.
\label{eq: parton_SLL}
\end{equation}
The $\sim$ sign in the above equation indicates that the states on either side of the sign differ in details of how the projection to the LLL is carried out. We anticipate that the universality class of the topological phase that the wave function belongs to is insensitive to such details~\cite{Balram15a, Balram16b, footnote_projection_details}. A nice feature of the wave function given on the rightmost side of Eq.~\eqref{eq: parton_SLL} is that it can be evaluated for large system sizes because the Jain states can be evaluated for a large number of electrons using the Jain-Kamilla projection~\cite{Jain97, Davenport12, Balram15}. 
	
An intuitive way to think about why the wave functions given in Eq.~\eqref{eq: parton_SLL} could be viable in the SLL is as follows. The particle-hole symmetry relating fermionic states at $\nu$ and $1{-}\nu$ can be used to define, in an operational sense, a particle-hole transformation for bosons. This would relate a state at $\nu_{b}{=}\nu/(1{-}\nu)$ [wave function $\Psi_{\nu}/\Phi_{1}$] with that at $\nu_{b}{=}(1{-}\nu)/\nu$ [wave function $\Psi_{1{-}\nu}/\Phi_{1}$]. In particular, using the fermionic $\nu{=}1/3$ Laughlin state~\cite{Laughlin83} we can relate the bosonic states at $\nu_{b}{=}1/2$ [parton $11$ which is the relevant vortex-attachment factor of the Jain states that are stabilized in the LLL] and $\nu_{b}{=}2$ [parton $\bar{2}1$ which is part of the correlation factors for the proposed parton states to capture the SLL FQHE] using the above transformation. The bosonic $\nu_{b}{=}1/2$ Laughlin state has the property that it vanishes as $r^{2}$ when two particles separated by a distance $r$ are brought close to each other, and thus builds good short-range correlations which are ideal to stabilize FQHE states in the LLL. On the other hand, the $\nu_{b}{=}{2}$ bosonic wave function $\Psi^{\rm Jain}_{2/3}/\Phi_{1}$ does not vanish when two particles separated by $r$ are brought to the same point, and thus allows particles to come close to each other even at short distances. This factor can build good correlations for SLL states since the Coulomb interaction in the SLL is less repulsive at short distances than the Coulomb interaction in the LLL, and thus particles in the SLL can come closer to each other as compared to particles in the LLL. We note that the $\nu_{b}{=}{2}$ bosonic wave function $\Psi^{\rm Jain}_{2/3}/\Phi_{1}$ has a low overlap with the exact ground states of interactions such as the LLL Coulomb and $V_{0}$-Haldane pseudopotential~\cite{Haldane83}, which are dominated by their short-range part (overlaps between the exact ground states of these short-range dominated interactions and the $\nu_{b}{=}{2}$ bosonic wave function $\Psi^{\rm Jain}_{2/3}/\Phi_{1}$ are presented in the Appendix).
	
The $p{=}1$ primary sequence of Eq.~\eqref{eq: parton_SLL} results in states at $\nu{=}2/3,~1/2$, and $6/13$ for $n{=}1,~2$, and $3$. FQHE has been observed at all three fractions in the SLL~\cite{Willett87, Kumar10}, and the corresponding wave functions have shown to provide a good representation of the exact SLL Coulomb ground states~\cite{Ambrumenil88, Balram13b, Kusmierz18, Balram18, Balram18a}. The $p{=}2$ secondary sequence of Eq.~\eqref{eq: parton_SLL} results in states at $\nu{=}1/2,~2/5$, and $3/8$ for $n{=}1,~2$, and $3$ and FQHE has been observed at these three fractions also in the SLL~\cite{Willett87, Xia04, Pan08, Choi08, Kumar10, Zhang12}. Furthermore, the corresponding wave functions here too are viable representations of the exact SLL Coulomb ground states~\cite{Balram18, Balram19, Balram21}. We note here that the $n{=}2$ member of the secondary sequence, that occurs at $\nu{=}2/5$, lies in the same universality class as the particle-hole conjugate of the 3-cluster Read-Rezayi state (aRR3)~\cite{Read99, Balram19}.
	
An enticing feature of the parton sequence of Eq.~\eqref{eq: parton_SLL} is that the strength of the FQHE states in the SLL is as per the order in which the fractions appear in the sequence. Moreover, aside from the fractions that lie in the secondary $n/(4n{\pm}1)$ Jain sequence, these are the \emph{only} fractions observed in the SLL. The LLL and SLL Coulomb ground states falling in the $n/(4n{\pm}1)$ sequence are expected to be analogous to each other and are well-represented by the Jain states~\cite{Balram21}. Following along the primary and secondary sequence of Eq.~\eqref{eq: parton_SLL}, the fractions at which FQHE is expected to occur next in the SLL would be $4/9$ and $4/11$ respectively. Signs of FQHE have been observed at $2{+}4/9$~\cite{Pan08, Choi08, Reichl14}, and the authors of Ref.~\cite{Balram20b} showed that the corresponding parton state is feasible in the SLL. In this work, we'll consider FQHE at $\nu{=}2{+}4/11$ where as yet no signs of FQHE have been reported. Nevertheless, with improving sample quality~\cite{Chung21}, it is plausible that FQHE could be realized at this fraction of $26/11$.

\subsection{Candidate states for $\nu{=}2{+}4/11$}
\label{subsec: parton_4_11}
Specifically, we consider the parton state denoted as $\bar{4}\bar{2}^{2}1^{4}$ and described by the wave function
\begin{equation}
\Psi^{\bar{4}\bar{2}^{2}1^{4}}_{4/11} = \mathcal{P}_{\rm LLL} [\Phi_{4}]^{*}[\Phi^{2}_{2}]^{*}\Phi^{4}_{1} \sim \frac{\Psi^{\rm Jain}_{4/7}[\Psi^{\rm Jain}_{2/3}]^{2}}{\Phi^{2}_{1}}
\label{eq: parton_4_11_bar4bar2bar21111}
\end{equation}
to capture the ground state of the potential FQHE phase that can occur at $2{+}4/11$. This state has a Wen-Zee shift~\cite{Wen92} of $\mathcal{S}^{\bar{4}\bar{2}^{2}1^{4}}{=}{-}4$. The presence of a repeated factor of $\Phi_{2}$ makes this state non-Abelian~\cite{Wen91}. Aside from this state, there are two other potential candidate states at $4/11$ which have been proposed in the literature which we describe next.
	
Bonderson and Slingerland (BS)~\cite{Bonderson08} have proposed the following wave functions:
\begin{equation}
\Psi^{\rm BS}_{n/(3n-1)} = \mathcal{P}_{\rm LLL} {\rm Pf}\left( \frac{1}{z_{i}-z_{j}}\right) [\Phi_{n}]^{*}\Phi^{3}_{1} \sim \Psi^{\rm MR}_{1}\Psi^{\rm Jain}_{n/(2n-1)},
\label{eq: BS_wf}
\end{equation}
which can capture many of the FQHE ground states occurring in the SLL. Here, $\Psi^{\rm MR}_{1}$ is the $\nu{=}1$ bosonic Moore-Read Pfaffian state~\cite{Moore91} and this factor makes the BS states non-Abelian. The $n{=}1$ member of the BS sequence is identical to the fermionic $\nu{=}1/2$ Moore-Read state~\cite{Moore91} which is a viable candidate for $5/2$. The $n{=}2$ member produces a candidate state at $2/5$ which has been shown to be competitive~\cite{Bonderson12} with the aRR3 state~\cite{Read99}, however, recent studies favor the aRR3 state~\cite{Wojs09,Zhu15,Mong15,Pakrouski16} at $12/5$. The $n{=}3$ member results in a state at $3/8$ and is a feasible candidate to describe the ground state at the experimentally observed $2{+}3/8$ fraction~\cite{Hutasoit16}. We will consider the $n{=}4$ member in this work which occurs at $4/11$. The $4/11$ BS state has a Wen-Zee shift of $\mathcal{S}^{\rm BS}{=}0$. The fraction $6/13$, where FQHE has been observed in the SLL~\cite{Kumar10}, is not a member of the BS sequence of Eq.~\eqref{eq: BS_wf}.
	
For completeness, we also consider the Abelian parton state denoted as $4\bar{2}1^{3}$ and described by the wave function
\begin{equation}
\Psi^{4\bar{2}1^{3}}_{4/11} = \mathcal{P}_{\rm LLL} \Phi_{4}[\Phi_{2}]^{*}\Phi^{3}_{1} \sim \frac{\Psi^{\rm Jain}_{4/9}\Psi^{\rm Jain}_{2/3}}{\Phi_{1}}.
\label{eq: parton_4_11_4bar2111}
\end{equation}
This wave function has been shown to give a good representation of the experimentally observed $4/11$ state in the LLL~\cite{Samkharadze15b, Pan15, Balram21c}. On the spherical geometry, this state occurs at the Wen-Zee shift of $\mathcal{S}^{4\bar{2}1^{3}}{=}5$.

\section{Results}
\label{sec: results}
All our numerical calculations are carried out on the Haldane sphere~\cite{Haldane83}. In this geometry, $N$ electrons move on a spherical surface of radius $R{=}\sqrt{Q}\ell$, where $\ell{=}\sqrt{\hbar c/(eB)}$ is the magnetic length at field $B$, in the presence of a radial magnetic flux of strength $2Qhc/e$ ($2Q$ is an integer) generated by a monopole placed at the center of the sphere. The flux-particle relationship for a state on the sphere is written as $2Q{=}\nu^{{-}1}N{-}\mathcal{S}$, where the Wen-Zee shift $\mathcal{S}$ is a quantum number characterizing the state~\cite{Wen92}. 

The IQHE state of $n$ filled LLs occurs when $2Q{=}N/n{-}n$ and requires $N$ to be divisible by $n$ with $N{\geq}n^2$. Therefore, the $\bar{4}\bar{2}^{2}1^{4}$ state can be constructed on the sphere only when $N{=}16{+}4k$, where $k$ is a nonnegative integer. In the spherical geometry, the smallest system for which the $\bar{4}\bar{2}^{2}1^{4}$ state can be realized has $N{=}16$ electrons. This system has a Hilbert space of over 28 billion and thus is well beyond the reach of exact diagonalization. Thus, we cannot compare the $\bar{4}\bar{2}^{2}1^{4}$ ansatz against exact results. Therefore, we can only study the competition between the candidate states proposed in the previous section. We will compare the states in the ideal SLL setting and neglect the effects of LL mixing, the finite width of the semiconductor quantum well, screening by gates, and disorder. Throughout this work, we will carry out LLL projection of the candidate wave functions using the Jain-Kamilla method~\cite{Jain97, Davenport12, Balram15}.

\subsection{Effective interaction}
\label{subsec: eff_int}
The wave functions we presented in Sec.~\ref{subsec: parton_4_11} can be readily evaluated only in the LLL. Haldane showed~\cite{Haldane83} that the problem of interacting electrons in any LL is equivalent to electrons confined to the LLL and interacting via an effective interaction. This effective interaction has the same Haldane pseudopotentials $V_{m}$ (which is the energy cost of placing two electrons in a relative angular momentum $m$ state) in the LLL as that of the Coulomb interaction in the given LL. To simulate the physics of the SLL in the LLL, we use the following form of the effective interaction~\cite{Shi08, Balram13b},
\begin{equation}
V^{\rm eff}(r) = \frac{B_{1}}{r}+\frac{B_{3}}{\sqrt{r^6+1}}+\frac{B_{5}}{\sqrt{r^{10}+10}}+\sum_{l=0}^{l=6} C_{l}r^{2l}e^{-r^2},
\label{eq: effective interaction}
\end{equation}
that has been used previously but we shall determine the coefficients $B_{l}$ and $C_{l}$ in a way that is different from the previous approaches. Since the different candidate states used here could be energetically close it is important to model the effective interaction accurately. Previously, effective interactions that fit only the odd pseudopotentials, i.e., which model the electron-electron interactions accurately, were deployed with the contribution of the constant positively charged background taken to be the same as in the LLL~\cite{Balram15c, Balram21}. Using the effective interaction obtained by fitting only the odd pseudopotentials is fine when evaluating energy differences such as those that arise in neutral gap calculations since energy differences do not depend on the contribution of the background. However, per-particle energies of states are sensitive to the background contribution~\cite{Balram20b}. Thus, when evaluating the per-particle energies of competitive states it is important to construct an effective interaction that fits both the even and odd pseudopotentials and thereby captures the background contribution accurately.

%%%%%%%%%%%%%%%%%%%%%%%%%%%%%%%%%%%%%%%%%%%%%%%%%%%%%%%%%%%%%%%%%%%%%%%%%%
\begin{table}[H]
	\begin{center}	
		\begin{tabular}{|c | c |} 	
			
			\hline
			Coefficient & Value \\
			\hline
			$B_1$ & 1.0 \\ 
			\hline
			$B_3$ & 1.0 \\
			\hline
			$B_5$ & 2.25 \\
			\hline
			$C_0$ & 237.90227\\
			\hline
			$C_1$ & -1450.65196\\
			\hline
			$C_2$ & 1858.22519\\
			\hline
			$C_3$ & -869.27320\\
			\hline
			$C_4$ & 175.54086\\
			\hline
			$C_5$ & -15.40401\\
			\hline
			$C_6$ & 0.47624\\
			\hline
		\end{tabular}
	\end{center}
	\caption{ Coefficients of effective interaction given in Eq.~\eqref{eq: effective interaction}.}
	\label{tab: coefficients_eff_int}
\end{table}
%%%%%%%%%%%%%%%%%%%%%%%%%%%%%%%%%%%%%%%%%%%%%%%%%%%%%%%%%%%%%%%%%%%%%%%%%%

We determine the coefficients $C_{l}$ by fitting the effective interaction's pseudopotentials in the LLL to match the first seven SLL \emph{disk} pseudopotentials (including the first four even and the first three odd). This captures the short-range part of the interaction accurately. The coefficients $B_{l}$, as was also done previously, are chosen to match the long-range part of the Coulomb interaction in the SLL. The final set of coefficients obtained from our fitting is given in Table~\ref{tab: coefficients_eff_int}. Although the effective interaction's pseudopotentials are matched with the disk pseudopotentials, we shall use the interaction to evaluate energies in the spherical geometry. We chose to do the fitting to match the disk pseudopotentials as opposed to the spherical ones since the former does not depend on the system size while the latter does and thus we can use the same effective interaction for all systems. We will go to fairly large systems and evaluate the thermodynamic limit of the energies so we expect that our results are not very sensitive to the precise choice of pseudopotentials that the effective interaction is matched to. 

The total energy of a state is given by
\begin{equation}
E=E_{ee}+E_{eb}+E_{bb},
\label{eq: total energy}
\end{equation}
where $E_{ee}$, $E_{eb}$, and $E_{bb}$ are the electron-electron, electron-background, and background-background energies, respectively. For a given wave function, its $E_{ee}$ is calculated as the expectation value of the effective interaction given in Eq.~\eqref{eq: effective interaction} using the Metropolis-Monte Carlo method~\cite{Binder10}. The $E_{eb}$ and $E_{bb}$ are only a function of the density $\rho{=}N/(4\pi R^{2})$ and their net contribution for the effective interaction given in Eq.~\eqref{eq: effective interaction} is given by

\begin{eqnarray}
E_{eb}+E_{bb} &=& -\frac{N^{2}}{4} \int_0^{\pi} \sin(\theta)d\theta~V^{\rm eff} \left( 2 R \sin \left(\frac{\theta}{2} \right) \right)   \\
 &=& -\frac{N^2}{4 R^2} \bigg [ 2 B_1 R+ 2 B_3 R^2 \, _2F_1\left(\frac{1}{3},\frac{1}{2};\frac{4}{3};-64 R^6\right) \nonumber \\
    && +\sqrt{\frac{2}{5}} B_5 R^2 \, _2F_1\left(\frac{1}{5},\frac{1}{2};\frac{6}{5};-\frac{1}{5} \left(512 R^{10}\right)\right) \nonumber \\
    && +C_0 \left(\frac{1}{2}-\frac{1}{2} e^{-4 R^2}\right) + C_1 \left(\frac{1}{2}-\frac{1}{2} e^{-4 R^2} \left(4 R^2+1\right)\right) \nonumber \\
    && +C_2 \left(1-e^{-4 R^2} \left(8 R^4+4 R^2+1\right)\right) \nonumber \\
    && +C_3 \left(3-e^{-4 R^2} \left(32 R^6+24 R^4+12 R^2+3\right)\right) \nonumber \\
    && -\frac{1}{2} C_4 \left(\Gamma \left(5,4 R^2\right)-24\right)+ C_5 \left(60-\frac{\Gamma \left(6,4 R^2\right)}{2}\right) \nonumber \\
    && -\frac{1}{2} C_6 \left(\Gamma \left(7,4 R^2\right)-720\right) \bigg ] \nonumber,
\end{eqnarray}
where $r{=}2R\sin(\theta/2)$ is the chord distance with respect to which we evaluate energies on the sphere, $\Gamma (s,x)$ is the incomplete gamma function, $_2F_1(a,b;c;z)$ is the Gaussian hypergeometric function~\cite{Arfken2013871}, and the coefficients $B_{l}$ and $C_{l}$ are given in Table~\ref{tab: coefficients_eff_int}. 
	
%%%%%%%%%%%%%%%%%%%%%%%%%%%%%%%%%%%%%%%%%%%%%%%%%%%%%%%%%%%%%%%%%%%%%%%%%%
\begin{figure}[htpb]
\begin{center}
\includegraphics[width=0.5\textwidth,height=0.27\textwidth]{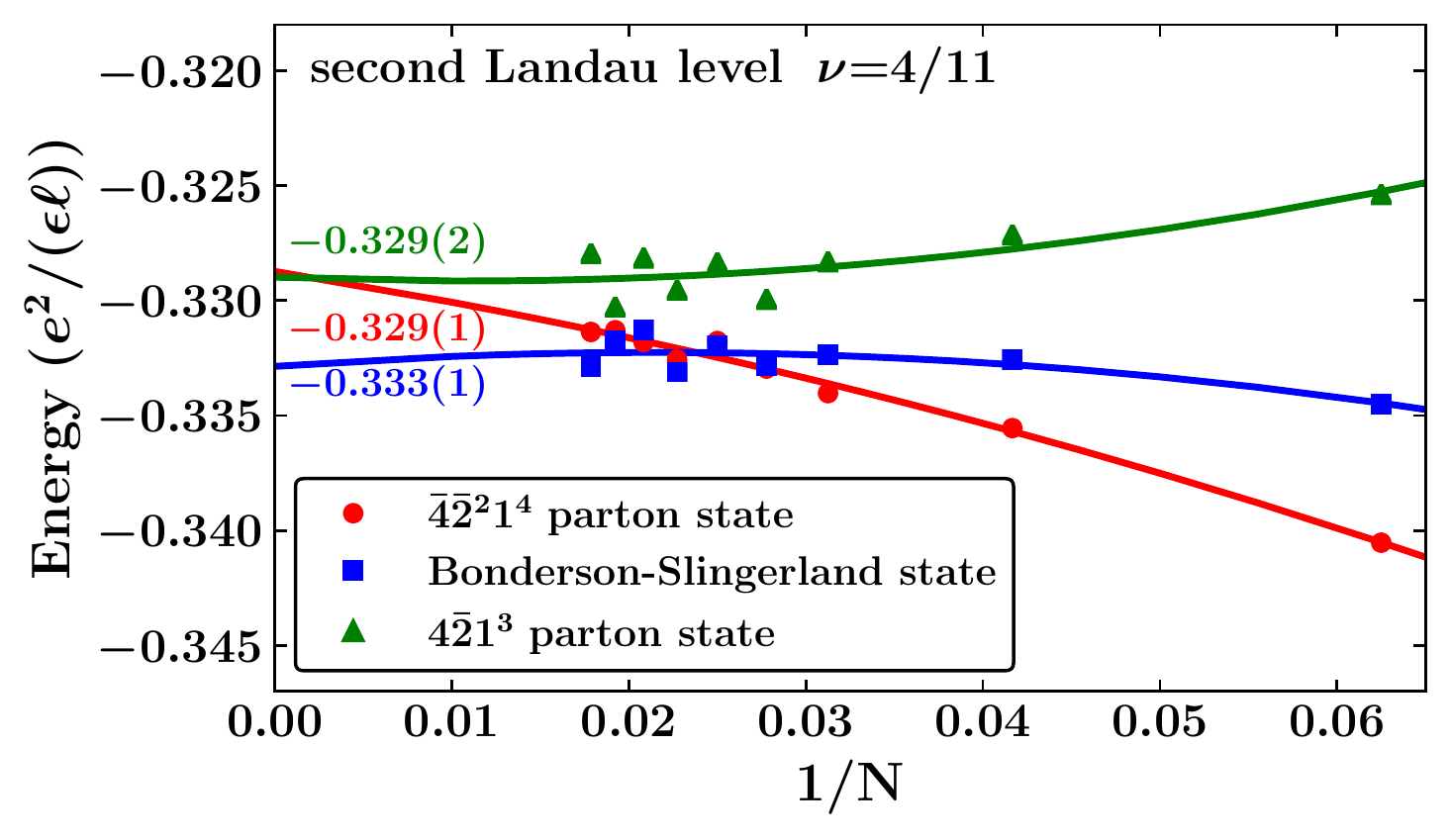}
\caption{Thermodynamic extrapolation of the per-particle density-corrected~\cite{Morf86} Coulomb energy of the $\bar{4}\bar{2}^{2}1^{4}$, Bonderson-Slingerland and $4 \bar{2} 1^3$ state of Eqs.~\eqref{eq: parton_4_11_bar4bar2bar21111},~\eqref{eq: BS_wf}, and ~\eqref{eq: parton_4_11_4bar2111}, respectively, in the second Landau level as a quadratic function of $1/N$, where $N$ is the number of electrons, evaluated in the spherical geometry. Data is shown for $N{=}16$ to $60$ electrons. The extrapolated energies, in Coulomb units of $e^2/(\epsilon \ell)$, where $\epsilon$ is the dielectric constant of the host material, are also shown on the plot. The parenthetical value in the extrapolated energies comes from the error in the quadratic fit.}
\label{fig: energies_pp_4_11_SLL}
\end{center}
\end{figure}
%%%%%%%%%%%%%%%%%%%%%%%%%%%%%%%%%%%%%%%%%%%%%%%%%%%%%%%%%%%%%%%%%%%%%%%%%%

In Fig.~\ref{fig: energies_pp_4_11_SLL} we show the total energies in the SLL of all three candidate states at $4/11$. As seen from the plot, all three states have comparable energies with the BS state having slightly lower energy than the other two (which are within error bars of each other). We note that several approximations have been made in our numerical computations such as ignoring effects of LL mixing, finite width of the quantum well, disorder, screening by gates, etc. which all need to be taken into account to do a realistic modeling of the experimental setting. Moreover, as stated above, the effective interaction of Eq.~\eqref{eq: effective interaction} approximates the ideal SLL Coulomb potential only in the large system limit. Furthermore, the projection of the parton as well as BS states is done in such a way that these are amenable to numerical evaluations. Owing to these reasons, as well as the close competition seen in our numerical results, we cannot definitively tell which state prevails in the thermodynamic limit at $2{+}4/11$. 

\begin{figure}[htpb]
\begin{center}
\includegraphics[width=0.47\textwidth,height=0.3\textwidth]{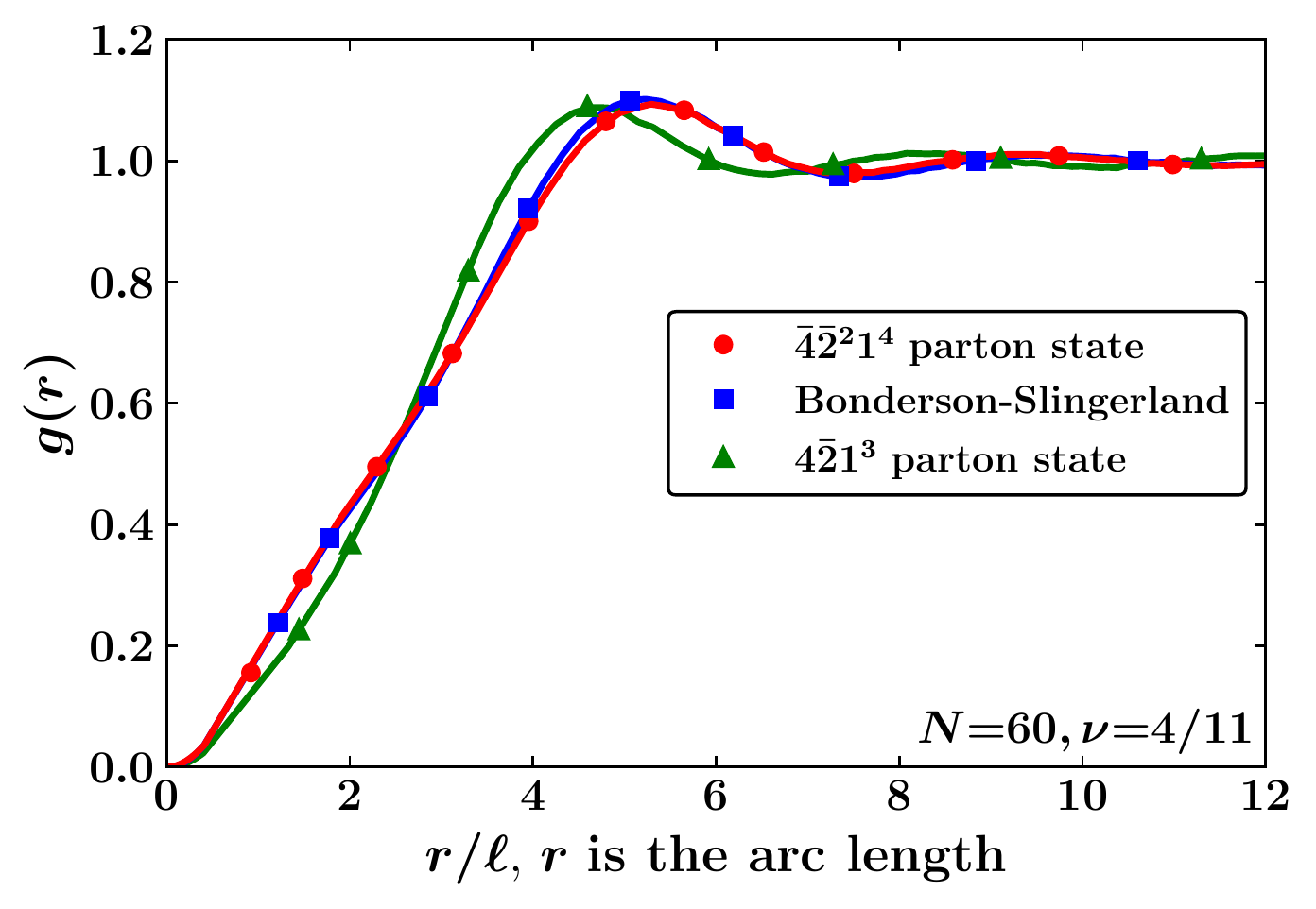}
\caption{The pair-correlation function $g(r)$ of the $\bar{4}\bar{2}^{2}1^{4}$, Bonderson-Slingerland and $4 \bar{2} 1^3$ states of Eq.~\eqref{eq: parton_4_11_bar4bar2bar21111}, \eqref{eq: BS_wf}, and \eqref{eq: parton_4_11_4bar2111}, respectively, as a function of the arc distance $r$ on the sphere for $N{=}60$ electrons. The value of $g(r)$ beyond $r{=}12\ell$ is near unity for all the states. }

\label{fig: pair_correlations_4_11_candidates}
\end{center}
\end{figure}

Next, to further characterize the various candidate states, we evaluate their pair-correlation function $g(r)$, which gives the probability of finding an electron a distance $r$ away from an electron that sits at the origin. In Fig.~\ref{fig: pair_correlations_4_11_candidates} we show the $g(r)$ of all three candidate states at $4/11$ for $N{=}60$ electrons. The $g(r)$ of all three states is similar to that of other incompressible quantum Hall fluids~\cite{Balram15b, Balram17} in that it has oscillations at intermediate distances and goes to the background density (unity, when appropriately normalized) at large distances. The presence of a ``shoulder"-like feature at short-to-intermediate distances in the $\bar{4}\bar{2}^{2}1^{4}$ and BS states, which typically exists in non-Abelian states~\cite{Read99, Hutasoit16, Balram20a}, is in accordance with these being non-Abelian, whereas the $\bar{4}\bar{2}1^{3}$ state shows no such feature which is consistent with it being Abelian. The $g(r)$ of the $\bar{4}\bar{2}^21^4$ and BS states are quite close to each other which further reinforces the point that these states are closely competing with each other. Although the $g(r)$ of the $4\bar{2}1^3$ states is different from that of the $\bar{4}\bar{2}^21^4$ and BS states, it is still competitive with them as seen from the energies shown in Fig.~\ref{fig: energies_pp_4_11_SLL}. 
	
%%%%%%%%%%%%%%%%%%%%%%%%%%%%%%%%%%%%%%%%%%%%%%%%%%%%%%%%%%%%%%%%%%%%%%%%%%
%%%%%%%%%%%%%%%%%%%%%%%%%%%%%%%%%%%%%%%%%%%%%%%%%%%%%%%%%%%%%%%%%%%%%%%%%%
	
Finally, we briefly discuss the nature of the neutral excitations of the $\bar{4}\bar{2}^{2}1^{4}$ state. The lowest-lying neutral excitation is expected to be obtained by creating a particle-hole pair in the $\Phi_{\bar{4}}$ factor since the corresponding parton carries the smallest charge of magnitude $e/11$. This collective mode is described by the wave function $\Psi^{{\rm mode-a}, L}_{4/11}{\sim}\Psi^{{\rm CFE}, L}_{4/7} [\Psi^{\rm Jain}_{2/3}]^{2}/\Phi^{2}_{1}$, where $\Psi^{{\rm CFE}, L}_{n/(2n\pm 1)}$ is the wave function of the CF-exciton (CFE) state at $\nu{=}n/(2n{\pm}1)$~\cite{Kamilla96b} with orbital angular momentum $L{=}2,3,{\cdots},(N/n{-}n){+}2n{-}1$~\cite{Balram16d}. Recently, it has been proposed that the parton theory can also describe certain very-high-energy collective modes of FQHE phases~\cite{Balram21d}. The parton ansatz of Eq.~\eqref{eq: parton_4_11_bar4bar2bar21111} readily suggests the construction of a high-energy neutral collective mode of the $\bar{4}\bar{2}^{2}1^{4}$ state that also starts from $L{=}2$. This high-energy mode can be obtained by creating a particle-hole pair in the $\Phi_{\bar{2}}$ factor; the corresponding parton carries a charge of magnitude $2e/11$. This excitation is described by the wave function $\Psi^{{\rm mode-b},L}_{4/11} {\sim}\Psi^{\rm Jain}_{4/7}\Psi^{\rm Jain}_{2/3} \Psi^{{\rm CFE},L}_{2/3}/\Phi^{2}_{1}$. Previous work~\cite{Balram21d} suggests that the particle-hole pair excitation in the $\Phi_{1}$ factor is likely projected out by $\mathcal{P}_{\rm LLL}$. The two modes $\Psi^{{\rm mode-a}, L}_{4/11}$ and $\Psi^{{\rm mode-b}, L}_{4/11}$ carry the same chiralities~\cite{Liou19} since the partons supporting them sense effective magnetic fields in the same directions~\cite{Balram21d}. In contrast, although the $4\bar{2}1^{3}$ and BS states also support two neutral collective modes starting from $L{=}2$, the two modes in both these states carry opposite chiralities. If FQHE is eventually observed at $\nu{=}2{+}4/11$ it would be worth studying the dispersion of these modes from which useful experimentally measurable quantities like the transport and roton gaps can be inferred.
	
\section{Discussion}
\label{sec: discussion}
We now discuss experimentally testable properties of the $\bar{4}\bar{2}^{2}1^{4}$ state that can help identify its underlying topological structure and distinguish it from other candidate states. The fundamental or smallest charged quasihole is obtained by creating a particle in the $\Phi_{\bar{4}}$ factor. These quasiholes are Abelian and carry a charge of $e/11$. Abelian quasiholes with larger magnitude charge can be produced by creating a hole in the $\Phi_{1}$ factor which produces excitation of charge $4e/11$. Excitations in the repeated factor of $\Phi_{\bar{2}}$ carry a charge of magnitude $2e/11$ and possess non-Abelian braid statistics. The BS state also supports non-Abelian anyons while the $4\bar{2}1^{3}$ state has only Abelian anyons~\cite{Balram21a}. 
	
The three states can be experimentally distinguished by thermal Hall measurements. In the recent past, heat transport measurements have been carried out in the SLL~\cite{Banerjee17b}. Assuming a complete equilibration of the edge states, the thermal Hall conductance $\kappa_{xy}$ of the $\bar{4}\bar{2}^{2}1^{4}$ state, at temperatures much lower than the gap, is $\kappa_{xy}{=}{-}7/2[\pi^2 k^{2}_{B}/(3h)]T$~\cite{Balram19} [Note that the two lowest filled LLs of spin up and spin down provide an additional contribution of $2[\pi^2 k^{2}_{B}/(3h)]T$ to $\kappa_{xy}$]. In contrast, the $4/11$ BS state has $\kappa^{\rm BS}_{xy}{=}{-}3/2[\pi^2 k^{2}_{B}/(3h)]T$~\cite{Hutasoit16} while the $4\bar{2}1^{3}$ state has $\kappa^{4\bar{2}1^{3}}_{xy}{=}3[\pi^2 k^{2}_{B}/(3h)]T$~\cite{Balram21a}. 
	
Another topological quantity that has not yet been experimentally accessed but can tell the three states apart is the Hall viscosity~\cite{Avron95}. For states at $\nu{=}4/11$, the Hall viscosity $\eta_{H}$ is expected to be quantized~\cite{Read09} as $\eta_{H}{=}(4/11)\hbar \mathcal{S}/(8\pi\ell^{2})$, where $\mathcal{S}$ is the shift of the state. Therefore, for the $\bar{4}\bar{2}^{2}1^{4}$ state, $\eta_{H}{=}(4/11)\hbar ({-}4)/(8\pi\ell^{2})$, for the $4\bar{2}1^{3}$ state,  $\eta^{4\bar{2}1^{3}}_{H}{=}(4/11)\hbar (4)/(8\pi\ell^{2})$ while the Hall viscosity vanishes for the $4/11$ BS state. These properties of the candidate states at $4/11$ are summarized in Table~\ref{tab:4_11}.	
	
%%%%%%%%%%%%%%%%%%%%%%%%
 \begin{table}[h]
		\begin{center}
			\begin{tabular} { | c | c | c | c |}
				\hline
				\multicolumn{4}{|c|}{Candidate states at $\nu$ = 4/11}\\
				\hline
				State &$\mathcal{S}$ & $\kappa_{xy}~[\pi^2 k_{\rm B}^2 /(3h)]T$ & Abelian/non-Abelian\\
				\hline
				$4\bar{2}1^{3}$ [Eq.~\eqref{eq: parton_4_11_4bar2111}] & $5$ & $3$ & Abelian \\
				\hline
				4/11 BS [Eq.~\eqref{eq: BS_wf}] & $0$ & $-3/2$ & non-Abelian \\
				\hline
				$\bar{4}\bar{2}^{2}1^{4}$ [Eq.~\eqref{eq: parton_4_11_bar4bar2bar21111}] & $-4$ & $-7/2$ & non-Abelian \\
				\hline
			\end{tabular}
		\end{center}
		\caption{\label{tab:4_11} Candidate states at $\nu{=}4/11$ with their Wen-Zee shift $\mathcal{S}$ on the sphere, thermal Hall conductance $\kappa_{xy}$ in units of $[\pi^2 k_{\rm B}^2 /(3h)]T$ and nature of excitations (Abelian/non-Abelian). }
	\end{table}
%%%%%%%%%%%%%%%%%%%%%%%%

Owing to the presence of the $\Phi_{\bar{n}}$ factors, all three states support upstream modes. The tunneling exponent of quasiparticles across the edge, which is an experimentally accessible quantity, for an FQHE state that does not have copropagating edge modes depends on the details of the interaction between the edge channels~\cite{Faugno19}. Thus, we expect that the tunneling exponents of none of the states are quantized to a universal value.

To summarize, we considered FQHE in the SLL at filling fraction $4/11$, where although a gapped state has not yet been observed experimentally, the parton sequence that captures other experimentally observed states in the SLL naturally suggests that FQHE could occur at. We considered three candidate states for FQHE at this filing. Numerical calculations showed tight competition between all three states suggesting that the underlying topological order at $4/11$ in the SLL can likely be determined only experimentally. To this end, we proposed a few experimental measurements which can conclusively determine the topological order at this fraction.
	
\begin{acknowledgements} 
Computational portions of this research work were conducted using the Nandadevi supercomputer, which is maintained and supported by the Institute of Mathematical Science's High-Performance Computing Center. We thank the Science and Engineering Research Board (SERB) of the Department of Science and Technology (DST) for funding support via Start-up Grant No. SRG/2020/000154.
\end{acknowledgements}
	
\begin{appendices}
\section{Bosonic states derived from fermionic states by particle-hole conjugation}
\label{app: bosonic_nu_2}
In this Appendix, we discuss some of the bosonic states that can be obtained from fermionic states by the notional definition of particle-hole conjugation described in the main text. This recipe involves taking any known LLL fermionic state at filling factor $\nu$, particle-hole conjugating it, and dividing this wave function by a Jastrow factor to obtain a bosonic state at $\nu_{b}{=}(1{-}\nu)/\nu$. Starting with the $1/3$ Laughlin state and applying the above transformations, one ends up with a $\nu_{b}{=}2$ bosonic state. This $\nu_{b}{=}2$ bosonic state can also be represented by the $\bar{2}1$ parton state and is part of the correlation factors present in the parton sequence of Eq.~\eqref{eq: parton_SLL} proposed for the SLL.
		
In Table~\ref{tab:overlaps_n_0_LL_bar21_aLaughlin} we show overlaps of this $\nu_{b}{=}2$ bosonic state with the lowest Landau level Coulomb and the $V_{0}$ model interactions' ground states obtained by exact diagonalization in the spherical geometry. The overlaps for small systems are sizable but drop quickly as the system size is increased indicating that it is unlikely that this $\nu_{b}{=}2$ bosonic wave function gives a good microscopic representation of the true ground states for these interactions that are dominated by their short-range part. Nevertheless, for all the systems accessible to us using exact diagonalization, we do find that the ground state at the shift corresponding to this bosonic state is uniform and gapped. Thus, it is possible that although the $\bar{2}1$ wave function does not give a very accurate representation of the actual ground state for interactions dominated by their short-range part, it might be in the right topological phase.

To obtain the numbers shown in Table~\ref{tab:overlaps_n_0_LL_bar21_aLaughlin} we constructed the $\nu_{b}{=}2$ bosonic state by particle-hole conjugating the $1/3$ Laughlin state and dividing the resulting $2/3$ state by a Jastrow factor. Instead of using the particle-hole conjugate of the Laughlin state, one could also directly use the $2/3$ Jain state (equivalently the $\bar{2}1^{2}$ parton state). However, these details on which among the two microscopically different representations for $2/3$ we use do not matter for our conclusions since for the systems sizes shown in Table~\ref{tab:overlaps_n_0_LL_bar21_aLaughlin}, the particle-hole conjugate of the $1/3$ Laughlin state and the $2/3$ Jain state are expected to have near unit overlap~\cite{Balram21b}. 
		
%%%%%%%%%%%%%%%%%%%%		
  \begin{table}[h]
			\centering
			\begin{tabular}{|c|c|c|}
				\hline
				$N$ & $|\langle\Psi^{\rm Jain}_{2} | \Psi^{0{\rm LL}}_{2} \rangle|^{2}$ & $|\langle\Psi^{\rm Jain}_{2} | \Psi^{V_{0}}_{2} \rangle|^{2}$ \\ \hline
				6	&		0.749	&	0.749	\\ \hline
				8	&		0.901	&	0.901	\\ \hline
				10	&		0.609	&	0.585	\\ \hline
				12	&		0.653	&	0.656	\\ \hline
				14	&		0.512	&	0.498	\\ \hline
				16	&		0.480	&	0.452	\\ \hline
				\end{tabular}
			\caption{\label{tab:overlaps_n_0_LL_bar21_aLaughlin} 
				Squared overlaps of the $\nu_{b}{=}2$ Jain state (equivalently the $\bar{2}1$ state), $\Psi^{{\rm Jain}}_{2}$ evaluated as the particle-hole conjugate of the $1/3$ Laughlin over the $\nu{=}1$ state with the Coulomb ground state in the lowest Landau level and the $V_{0}$ Haldane pseudopotential interaction for $N$ electrons in the spherical geometry. }
		\end{table}
%%%%%%%%%%%%%%%%%%%%

Similarly, one can obtain bosonic states at $\nu_{b}{=}2p$ from particle-hole conjugation of the $1/(2p{+}1)$ Laughlin states~\cite{Laughlin83} and some of these were recently considered in Ref.~\cite{Yang21}. Furthermore, following the procedure outlined above, one can obtain bosonic states from other fermionic states such as the $n/(2pn{\pm}1)$ Jain states~\cite{Jain89}, the $n\bar{n}1^{2p{+}1}$ parton states~\cite{Balram20}, etc. No realistic interaction that conclusively stabilizes any of these bosonic states is known. Bosonic states that are obtained by dividing the analogous fermionic states by a Jastrow factor, such as the $\nu_{b}{=}1/2$ Laughlin state~\cite{Laughlin83}, the $\nu_{b}{=}2/3$ and $\nu_{b}{=}3/4$ Jain states~\cite{Jain89}, and the $\nu_{b}{=}1$ Moore-Read state~\cite{Moore91}, can be stabilized by interactions that have a dominant short-range part~\cite{Xie91, Regnault03, Nakajima03, Chang05b}. 
		
The electronic state at low filling factors $\nu{\lesssim}1/9$ is expected to be a Wigner crystal~\cite{Zuo20}. Assuming the notion of particle-hole conjugation described in the main text holds, for the bosonic states at $\nu{\gtrsim}8$ we expect a crystal state to prevail in analogy to the fermionic state at low fillings. 

\end{appendices}
		
\bibliography{biblio_fqhe}

\begin{thebibliography}{77}
\bibitem[{foo()}]{footnote_projection_details}
\bibinfo{note}{There are examples known where the details of projection can
  matter. For example, the $n\bar{n}1^{3}$ states proposed in
  Ref.~\cite{Balram20}, when projected as $\Phi_{1}^{2}\mathcal{P}_{\rm
  LLL}\Phi_{n}[\Phi_{n}]^{*}\Phi_{1}$ are exactly identical to the $1/3$
  Laughlin state $\Phi_{1}^{3}$ (as $\Phi_{1}$ is the unique LLL state at
  $\nu{=}1$, which is the filling factor of the $n\bar{n}1$ state that is being
  projected to the LLL), whereas when projected as $(\mathcal{P}_{\rm
  LLL}\Phi_{n}\Phi^{2}_{1})(\mathcal{P}_{\rm
  LLL}[\Phi_{n}]^{*}\Phi^{2}_{1})/\Phi_{1}{\sim} \Psi^{\rm
  Jain}_{n/(2n{+}1)}\Psi^{\rm Jain}_{n/(2n{-}1)}/\Phi_{1}$ results in a state
  that is topologically distinct from the $1/3$ Laughlin state~\cite{Balram20}.
  We anticipate that instances where $\Phi_{n}$ and $[\Phi_{n}]^{*}$ are
  grouped separately and projected results in a state that lies in the latter
  universality class that is distinct from the $1/3$ Laughlin state.}
\end{thebibliography}
\bibliographystyle{apsrev}
\end{document}